\documentstyle[12pt,epsfig]{article}

\large
\newcommand{\be}{\begin{eqnarray}}
\newcommand{\ee}{\end{eqnarray}}
\newcommand{\bea}{\left (\begin{array}{cc}}
\newcommand{\eea}{\right )\end{array}}

\newcommand{\mat}{\left ( \begin{array}{cc}}
\newcommand{\emat}{\end{array} \right )}
\newcommand{\matt}{\left ( \begin{array}{ccc}}
\newcommand{\ematt}{\end{array} \right )}
\newcommand{\matf}{\left ( \begin{array}{cccc}}
\newcommand{\ematf}{\end{array} \right )}
\newcommand{\vect}{\left ( \begin{array}{c}}
\newcommand{\evect}{\end{array} \right )}
\newcommand{\nn}{\nonumber}

\begin{document}
\setlength{\baselineskip}{17pt}
\vfill
\eject
\begin{flushright}
SUNY-NTG-01/10
\end{flushright}

\vskip 2.0cm
\centerline{\Large 
\bf Critical statistics in quantum chaos}
\vskip 0.4cm
\centerline{\Large \bf and Calogero-Sutherland model at finite temperature }

\vskip 0.8cm
\centerline{A.M. Garc\'{\i}a-Garc\'{\i}a and J.J.M. Verbaarschot}
\vskip 0.2cm
\centerline{\it
Department of Physics and Astronomy, SUNY, 
Stony Brook, New York 11794}
\vskip 1.5cm

\centerline{\bf Abstract}
\noindent
We investigate the spectral properties of a generalized GOE
(Gaussian Orthogonal Ensemble) capable of describing critical statistics. The 
joint distribution of eigenvalues of this model is expressed 
as the diagonal 
element of the density matrix of a gas of particles governed by the 
Calogero-Sutherland Hamiltonian (C-S). 
Taking advantage 
of the correspondence between C-S particles
 and eigenvalues, we show that the number variance 
of  our  random matrix model is asymptotically linear
with a slope depending on the parameters of the model. Such linear behavior 
is a signature of critical statistics. This random matrix model  
may be relevant for the description of spectral correlations of 
complex quantum systems with a self-similar/fractal Poincar\'e section of its
classical counterpart. This is shown in detail for  
two examples: the anisotropic Kepler problem 
and a kicked particle in a well potential. In both cases the
number variance and the $\Delta_3$-statistic is accurately described
by our analytical results.


\vfill

\eject

\section{Introduction}

Random matrix ensembles (RME)  are an invaluable
tool in describing  the level statistics of complex quantum systems. 
Typically, their range of 
applicability for disordered systems 
 is determined by the Thouless energy which, 
in the metallic phase,
is much larger than the average level spacing. In the neighborhood of a 
 localization-delocalization transition  
the Thouless energy is of the order of the average level spacing and the 
 wavefunctions become multifractal. The usual
random matrix ensembles are no longer applicable.
Recently, new random matrix ensembles 
\cite{kra1,Moshe,dittes,log2,mirlin,bog,us,bleck}
 depending on additional parameters have been proposed 
 to describe 
spectral correlations in this critical case.
These new models for critical statistics 
 have been successfully utilized 
 to describe the spectral correlations of
a disordered system at the Anderson transition in three dimensions
\cite{shk,nishigaki}, two dimensional Dirac fermions
in a random potential {\cite{caux}}, the quantum Hall transition
\cite{klesse} and of the QCD Dirac operator in a liquid of instantons
\cite{james,us}.
 
There are two different types of models for critical statistics.
In the first one, deviations from 
Wigner-Dyson statistics are obtained by  
adding a symmetry breaking term to the GUE {\cite{Moshe,us}}. 
The model is solved by mapping it to a non-interacting Fermi gas of 
 eigenvalues. The second one {\cite{log2}} makes use of  
soft confining potentials and 
is solved exactly by means of $q$-orthogonal polynomials.
Both models lead to the same spectral kernel for 
small deviations from the GUE. Based on this observation it was
conjectured \cite{kra1} that  critical statistics is universal.
However,
the origin of the critical kernel 
is different in both cases. In models based 
on a soft confining potential the 
critical kernel is obtained from a 
nontrivial unfolding. In 
models with an explicit symmetry breaking term, deviations from Wigner-Dyson
statistics arise because the long range correlations between the 
eigenvalues are
exponentially suppressed \cite{krav}. 
We remark that the extent of universality 
in critical statistics is still under debate.
 For instance, for chiral ensembles, the correlation functions of  
a model based on a symmetry breaking term \cite{us} and a model with a  
soft confining potential \cite{myself} are different. 
Only the former one reproduces critical statistics.
 
Critical random matrix models for orthogonal and symplectic ensembles 
have recently \cite{tsvlik} been reported in the literature. Tsvelik and
Kravtsov have obtained asymptotic expressions for the critical two
level spectral  function 
from a generalized ensemble of random banded matrices. An exact 
expression for the two level critical spectral 
function was conjectured in \cite{tsvlik} 
 for orthogonal and symplectic ensembles. In the context 
of the Anderson model, 
 a similar result  was conjectured by Nishigaki \cite{nishigaki}.

In order to describe the spectral correlations  of certain pseudo-integrable 
billiards with dynamics intermediate between chaotic and integrable, 
Bogomolny and coworkers \cite{bogo3} have 
 introduced a short range plasma model that interpolates between Poisson 
statistics and
Wigner-Dyson  statistics.
  The joint distribution of eigenvalues  in {\cite{bogo3}} is given by 
the classical Dyson gas with the logarithmic 
pairwise interaction restricted to a finite number $k$ of nearest neighbors. 
 Analytical solutions are available for general $k$ and symmetry 
 class.  It turns out that this short-range plasma model 
 reproduces the typical characteristics of critical statistics 
like a linear number variance, with a slope depending on $k$ and, 
asymptotically, an exponential decay of the nearest neighbor 
spacing distribution.
 However the two models models are not identical. In critical 
random matrix models based on a
 symmetry breaking term, the  joint distribution 
of eigenvalue  
 can be considered as an ensemble of free particles at finite temperature 
with a nontrivial statistical 
 interaction. The statistical interaction resembles
 the Vandermonde determinant, and the effect of finite temperature 
 is to suppress the correlations of distant eigenvalues. 
In \cite{bogo3} this suppression is abrupt, in contrast to critical 
statistics, where the effect 
 of the temperature is smooth.  For further details we refer to 
\cite{krav}.
 
In this paper we introduce a generalized GOE
 based on the addition of a symmetry breaking term to an invariant
Gaussian probability distribution along
 the lines of \cite{Moshe} for the GUE. We show that the joint eigenvalue
distribution of this model
 coincides with the diagonal element of the density matrix 
of a gas of  particles
governed by the Calogero-Sutherland (C-S) Hamiltonian. Using this 
identification
we calculate the asymptotic behavior of the number variance from the
susceptibility of the C-S partition function. Because the Itzykson-Zuber
integral for $\beta = 1$ is unknown, a direct calculation  of the
correlation functions is not possible. To obtain analytical results we
invoke the Kravtsov-Tsvelik conjecture, which states that the 
finite temperature modifications of the 
correlation functions arise only through the 
known finite temperature modifications of the kernel for $\beta =2$.
The validity of this conjecture is tested in two different ways. 
First, we show that it is in agreement with a conformal
calculation
of the asymptotic behavior of the two point correlation function. Second,
the asymptotic behavior of the number
variance according to the Kravtsov-Tsvelik conjecture agrees with the
behavior
of the susceptibility in the grand canonical ensemble. 
One of the main aims of this article is to show that critical statistics  
 describes the spectral
correlations of  time-reversal invariant quantum systems with a
corresponding classical  phase space that has a global self-similar, 
fractal structure. This is shown in two examples, 
a kicked particle in a 
potential well and the anisotropic Kepler problem, by comparing the
two-point level correlations with the analytical formula of our
critical Random Matrix Model.

The critical Random Matrix Model and its relations with the C-S model
is discussed in section 2. In section 3, we review the Kravtsov-Tsvelik
 conjecture for the density-density correlation function
of the C-S model
in the 
 low temperature limit. The validity of this conjecture is discussed in
section 4. In section 5 we show that the level correlations of a kicked
particle in a potential well and of the anisotropic Kepler problem are described
by the Kravtsov-Tsvelik conjecture. Concluding remarks are made in section 6.

\section{Definition of the Model}

A Random Matrix Model for Hermitian matrices with 
critical eigenvalue statistics was introduced in \cite{Moshe}. 
Although it is straightforward to generalize this model to the 
class of the Gaussian Orthogonal Ensemble, the absence of an explicit 
result for the  
integral over  orthogonal matrices makes its analysis far
more complicated. The model we study is defined by the joint probability
distribution
\be
\label{kl1}
P(S,b)=\int dM e^{-\frac{1}{2}{\rm Tr} SS^{T}} e^{-\frac{b}{2}
{\rm Tr}{[M,S][M,S]^{T}}}.
\ee
Here, the $N\times N$ matrices  $S$ and $M$ are real symmetric and
orthogonal, respectively, and
the integration measure $dM$ is the Haar measure.
{}From the invariance of $dM$ it follows that $P(S,b)$ is a
function of the eigenvalues of $S$ only. If the eigenvalues of $S$ are
denoted by $x_k$ the joint eigenvalue distribution is given by
\be
\label{kl}
\rho(x_1,\cdots,x_N}) = \Delta(\{x_k\}) 
\int dM e^{-\frac{1}{2}{(2b+1)\sum_k x_k^2 + 
b\sum_{k,l} M_{kl}^2 x_k x_l}.
\label{joint}
\ee
where the Vandermonde determinant is defined by
\be
\Delta(\{x_k\}) = \prod_{k<l} (x_k -x_l).
\label{vandermonde}
\ee
Let us now consider the harmonic oscillator Hamiltonian
\be
\hat H = -\nabla_S^2 + \frac 14 \omega^2 {\rm Tr} S^2
\label{hamiltonian}
\ee
where the Laplacian for symmetric matrices $S$ is given by
\be
\nabla_{S}^{2}=\sum_{i=1}^{N}\frac{\partial^{2}}{\partial S_{ii}^2}+
\frac{1}{2}\sum_{i<j}^{N}\frac{\partial^{2}}{\partial S_{ij}^2}.
\ee
This Hamiltonian is the sum of $N(N+1)/2$ independent harmonic oscillators
with imaginary time propagator given by \cite{zirnhaldane}
\be
\langle S |e^{-\tau \hat H } | S' \rangle = 
\left( \frac{\omega}{4\pi \sinh \omega 
\tau} \right )^{N(N+1)/4}
e^{-\frac{\omega}{4 \sinh \omega \tau}[({\rm Tr} S^2 + {\rm Tr} {S'}^2)
\cosh \omega \tau - 2 {\rm Tr} S S']}.
\label{propagator}
\ee
Since the Laplacian, $\nabla^2_S$, is the sum of a radial piece, depending
only on the eigenvalues of $S$, and an  angular piece, depending only on the
orthogonal matrix $M_S$ that diagonalizes $S$, 
\be 
\nabla^{2}_{S}=\frac{1}{\Delta(\{x_k\})}\sum_{i=1}^{N}\frac
{\partial}{\partial x_{i}}\Delta(\{x_k\})\frac
{\partial}{\partial x_{i}}+\nabla^{2}_{M_{S}},
\ee
the matrix element  $\langle S |e^{-\tau \hat H } | S' \rangle$ factorizes 
into a radial piece and an angular piece. After integration over the angular
degrees of freedom  and putting the eigenvalues of $S'$ equal to the 
eigenvalues of $S$ we  obtain 
\be
\langle x_1,\cdots , x_N |\Delta^{1/2}(\{x_k\})
e^{-\tau H_{\rm rad}} \Delta^{-1/2}(\{x_k\})  
|x_1,\cdots , x_N \rangle = 
C \int dM  e^{-\frac{\omega}{2 \sinh \omega \tau}[{\rm Tr} S^2 
\cosh \omega \tau -  {\rm Tr} S M S M^T]},\nn \\
\ee
where the integral over the angular matrix element 
has been  absorbed in the normalization constant $C$. 
If we make the identification
\be
\frac{\omega}{ \sinh \omega \tau} = 2b \quad {\rm and}
\quad \frac{\omega \cosh \omega \tau}{ \sinh \omega \tau} = 2b+1,
\label{beta}
\ee
the r.h.s. of this equation is exactly the joint probability 
distribution (\ref{joint}). 
We thus have shown that the joint probability
distribution of our model is given by  the diagonal matrix element of the
density matrix of 
the Hamiltonian
\be
\hat H =  \Delta^{1/2}(\{x_k\}) H_{\rm rad} 
\Delta^{-1/2}(\{x_k\}) .
\ee
Using the identity,
\be
 \Delta^{-1/2}(\{x_k\})  \sum_{i=1}^{N}
\frac{\partial}{\partial x_i} \Delta(\{x_k\})
\frac{\partial}{\partial x_i}
\Delta^{-1/2}(\{x_k\})  =\sum_{i=1}^{N}\frac
{\partial^2}{\partial x_i^2}
+\frac14 \sum_{k\ne l} \frac 1{(x_k-x_l)^2}, 
\ee
we find the Hamiltonian
\be
\label{er}
{\hat H}=-\sum_{j}\frac{\partial^2}{\partial
x_{j}^{2}}-
\frac 14 \sum_{i \neq j}\frac{1}{(x_{i}-x_{j})^2}+
\frac{\omega^{2}}{4}\sum_j x_{j}^{2}.
\label{csm}
\ee
This Hamiltonian corresponds to the Calogero-Sutherland model \cite{suth,calo}
\be
{\hat H_{C-S}}=-\sum_{j}\frac{\partial^2}{\partial
x_{j}^{2}}+{\frac{\lambda}2}(\frac{\lambda}2-1)\sum_{i \neq j}\frac{1}{(x_{i}-x_{j})^2}+ \frac{\omega^{2}}{4}\sum_j x_{j}^{2}.
\label{csm1}
\ee
with $\lambda =1 $ and fermionic boundary conditions.
 We have thus shown that the joint eigenvalue
distribution of the model (\ref{kl1}) is given by the diagonal matrix
elements of the $N$-particle density matrix of the Calogero-Sutherland 
model at an inverse temperature $\tau$ given by (\ref{beta}).

The normalized eigenfunctions of the Calogero-Sutherland Hamiltonian 
(\ref{csm}) can be
labeled in terms of the partitions  of integers 
denoted by $\kappa$ (see next section). 
If $\lambda_\kappa$ and $\Psi_\kappa(x_1,\cdots, x_N,\omega)$ 
are the eigenvalues
and eigenfunctions of the C-S Hamiltonian,
respectively, the joint eigenvalue probability distribution is given by
\be
\label{kas}
\rho(x_{1},\ldots,x_{N})=C'\sum_{\kappa}
e^{-\lambda_{\kappa}\tau}\Psi_{\kappa}(x,\omega)\Psi_{\kappa}(x,\omega)
\ee
where $C'$ is a constant and $x = x_1, \cdots ,x_n$. 
The $\Psi_\kappa(x,\omega)$ can be expressed  in
terms of the generalized Hermite polynomials \cite{forrest3}
\be
\Psi_k(x,\omega) = \frac 1{\sqrt{N_\kappa}}
e^{-\frac 14 \omega \sum_k x_k^2} \Delta^{1/2}(\{x_l\})
H_k(x\sqrt{\omega/2}, 2)
\ee
where $N_{\kappa}$
is a normalization constant
and the eigenvalue is given by
\be
\lambda_\kappa = \omega |\kappa|.
\ee
We point out that the above relation between the Euclidean propagator
in symmetric spaces and the C-S model at finite temperature can be
extended to all nine other symmetry classes  in the Cartan 
classification of large families of symmetric spaces \cite{class}. 
In essence, the radial part of the Laplacian 
 in the symmetric space corresponds to a C-S type Hamiltonian.
For a classification of C-S Hamiltonians based
on the symmetry  class we refer to {\cite{pomel}}.

Finally,  let us mention that the interpolating role
(between RMT and the  Poisson ensemble) of $b$ can be inferred
directly from (\ref{kl1}).
 Using the invariance
of the measure, the integral over $M$ can be replaced by an integral
over the eigenvalues of $M$. For $b \rightarrow \infty$, this partition
function is dominated by matrices
 $S$ that commute with  arbitrary
diagonal orthogonal matrices. This set of matrices is
the ensemble of diagonal symmetric
matrices also known as the Poisson ensemble with uncorrelated eigenvalues.
Critical statistics is obtained in the thermodynamic
limit if the parameter $b$ is scaled as
\be
\label{b}
b = h^2N^2.
\ee
 Wigner-Dyson statistics is found for a weaker $N$-dependence of $b$
whereas a stronger $N$-dependence leads to Poisson statistics. This 
transition 
can also be understood in terms of the C-S model
 at finite temperature. At zero temperature, the probability density 
 of the ground state of the C-S model (\ref{csm}) coincides with the
joint probability distribution of the Gaussian Orthogonal Ensemble.
In the high temperature  limit, $\tau \rightarrow 0$,   
the positions of the particles become uncorrelated and the 
statistics of the associated matrix model is Poisson. 

To recapitulate,
we have traded the
problem of performing an integral over the orthogonal group by the
physical task of finding the diagonal element of the density matrix
of an ensemble of  particles governed by the C-S Hamiltonian.

\subsection{ Excited states and Zonal Polynomials}

In this section we discuss explicit solutions of the 
 excited  eigenfunctions of
the C-S Hamiltonian (\ref{csm}) and argue to what extent they 
are useful for the evaluation of correlation functions from
the joint eigenvalue distribution (\ref{kl}).

The probability density of the ground state of the C-S model 
(\ref{csm}) coincides with the joint probability
distribution of the Gaussian Orthogonal Ensemble
\cite{suth}. This observation
together with the conjecture of the solvability of the C-S model
was already made in the pioneering articles of Calogero \cite{calo} and
Sutherland \cite{suth}. Later, Sutherland \cite{suth1}
obtained a non-orthogonal set of solutions. The problem
of finding a set of orthogonal solutions for these excited states was recently
solved by Forrester, Ha and Serban \cite{forrest,ha,ser} who expressed
the wavefunctions of the excited states in terms of
the symmetric Jack polynomials \cite{stan}. For the special values
 of the coupling constant related to GOE and GSE
the Jack polynomials have a geometrical interpretation
and are usually called zonal
polynomials \cite{muir}. Unfortunately, there is no
closed formula neither for the Jack nor
for the zonal polynomials \cite{muir}. Since explicit calculations rely
on recurrence relations, numerical work is needed
to evaluate polynomials of high degree.
In our case, due to the harmonic potential, the excites states are given by
the generalized Hermite (or Hidden-Jack) polynomials \cite{ujino} which
can be expressed in terms of Jack polynomials \cite{ujino,forrest3}.

The generalized Hermite polynomials
in (\ref{kas}) can be expressed in terms
of zonal polynomials  by means of a Mehler type formula \cite{forrest3}.
\be
\label{gh1}
\rho(x_1,\ldots,x_{N}) \propto
e^{-\frac 12 \omega \sum_k x_k^2} \Delta(\{x_k\})
\sum_\kappa \frac{e^{-\lambda_{\kappa}\tau}}{N_\kappa} 
H_\kappa(x\sqrt{\omega/2}, 2)
H_\kappa(x\sqrt{\omega/2}, 2)
\nonumber \\ \propto
\Delta(\{x_k\}  )e^{-\frac 12 \omega \coth(\tau \omega)
\sum_{i}x_{i}^{2}}\sum_{\kappa}\frac{1}{|\kappa|!}
\frac{C^{(2)}_{\kappa}(x\sqrt{2\omega}/\sqrt{1-e^{-2\omega \tau}}
)C^{(2)}_\kappa(x \sqrt{\omega/2}e^{-\tau \omega}/\sqrt
{1-e^{-2\omega \tau}} )}
{C^{(2)}_{\kappa}(1^{N})}
\ee
 where the $C^{(2)}_\kappa(x)$ are the symmetric Jack polynomials  as 
 defined in \cite{forrest3},
$x=x_{1},\ldots,x_{n}$ and $\kappa$ labels the partitions of 
the integers (there
 is a polynomial for each partition) and the sum runs over
all the partitions. Furthermore,
 $\tau$ and $\omega$ are related to $b$ through equation (\ref{beta}).

The kernel
\be
F_0(x,y) \equiv \sum_{\kappa}\frac{1}{|k|!}
\frac{C_{\kappa}^{(2)}(2x)C_{\kappa}^{(2)}(y)}{C_{\kappa}^{(2)}(1^N)}
\ee
has been studied extensively \cite{muir,forrest3}. 
Using the result for the $\tau \to 0$ limit of the kernel
\cite{forrest3},
\be
F_{0}(x/\sqrt{\tau},y/\sqrt{\tau}) 
\propto \frac{1}{\sqrt{\Delta(\{x_k\})\Delta(\{y_k\})}}\prod_{j=1}^{N}
e^{\tau x_{j}y_{j}/2}
\ee
one easily shows that the eigenvalues of our model are uncorrelated in
the high temperature limit.

In the zero temperature limit (GOE), 
 the joint distribution can be represented as a quaternionic determinant
and the integrations can be performed by means of a "kernel relation". 
At nonzero temperatures, the kernel $F_0(x,y)$ satisfies the "kernel relation" 
\cite{forrest3},
\be 
\int d\mu(y)F_{0}(2y,z)F_{0}(2y,x) 
\propto F_{0}(2z,x)e^{-\frac 12 \omega \sum_{i}(z^{2}_i+x_{i}^{2}}),
\ee  
where 
$d\mu(y)=\prod_{i=1}^{N}e^{-\frac{\omega}{2}y_{i}^{2}}\Delta(\{y_k\})
dy_{1}\ldots dy_{N}$, but it is not known\footnote{Some 
interesting results for
for small values of $N$ were obtained in \cite{kohler}.}
 whether the joint eigenvalue distribution
can be expressed in terms of quaternionic determinants.     
Further progress could rely on exploiting
 the orthogonality relations  \cite{muir} verified 
by the Zonal polynomials. 
 If proceeding so, the partition function associated with (\ref{kl1}) 
can be evaluated exactly.
As expected, it coincides with the one encountered for particles 
obeying fractional statistics in one dimension \cite{haldane}. 
Because of technical problems, the 
same strategy fails 
for the spectral density and higher order spectral correlation functions.

For the special case of $N=2$, an explicit 
calculation of the joint distribution of eigenvalues (\ref{kl}) 
coincides with the result obtained by using Zonal polynomials 
 techniques \cite{yan}.
\be
\rho(x_{1},x_{2})=\frac{1}{2\pi}e^{-\frac{1}{2}(x_{1}^{2}+x_{2}^{2})}I_{0}
(\frac b2(x_{1}-x_{2})^{2})e^{-\frac b2 (x_{1}-x_{2})^{2}}|x_1-x_2| 
\ee 
where $I_{0}$ is the Bessel function of complex argument.

In conclusion, we have expressed the
joint distribution for the eigenvalues 
of our model in terms of Jack polynomials but we have not succeeded to
derive explicit expressions for the correlation functions.

  \section{Critical spectral kernel and density-density
correlations of the C-S model at finite temperature}

We recall that the two level spectral
function of our model is identical to the density-density
correlation function of the C-S model (\ref{csm1}) for $\lambda = 1$.
In \cite{tsvlik} it was conjectured that
the low temperature limit  of  the 
connected density-density correlation function 
of the C-S model at $\lambda = 1$ is given by,
\be
\label{ic}
\langle \rho(x)\rho(0)\rangle_{T}-\langle \rho(x)\rangle_{T}
\langle \rho(0)\rangle_{T}
=R_{2,c}^T(x,0)= - {\bar K_T}^{2}
(x,0)-\left(\frac{d}{dx}{\bar K_T}(x,0)\right)\int_x^{\infty}{\bar K_T}(t,0)
 \ee
where 
\be
\label{ow}
 \bar K_T(x,0)=T\frac{\sin(\pi x)}{\sinh(\pi xT)}
\ee
is the kernel of the C-S model (\ref{csm1}) for $\lambda = 2$. The temperature 
$T=\frac{\pi h}{2}$ and $h$ is related to $\tau$ and $\omega$ through 
 (\ref{b}) and (\ref{beta}). This result is valid in a normalization such 
 that the average density of the particles is equal to unity.

 The idea is that,
 based on the Luttinger liquid nature of the C-S model \cite{kawa},
 the known relation between the density-density correlation of
the C-S for $\lambda=1$ at zero temperature
and the spectral correlations of the GOE can be extended
to finite low temperature. One simply replaces  
the kernel at zero temperature, which physically corresponds
to free fermions  
for all invariant RMT ensembles\footnote{For $\lambda \ne 2$ the
particles obey exclusion statistics},  by its finite
temperature  analogue \cite{Moshe} given by,
\be
\label{ui}
K^T(x, y)=\sum_n \frac{\psi_n(x) \psi^{\dagger}_n(y)}{1 + z^{-1} e^{\tau E_n}}
\ee
where $\psi_n$ are the single particle wave functions for free fermions 
and $E_n = \omega n$.
The fugacity $z$ for the free fermion distribution 
 in (\ref{ui}) is
determined by the total number of
particles
through,
\be
N = \int dx \rho(x) = \sum_{n=0}^{\infty}\frac{1}{{1+z^{-1}
e^{\omega \tau n}}}.
\ee
where $\rho(x)=K^{T}(x,x)$ is the average spectral density.
In the low temperature limit,
$\omega \tau \ll 1$,  we find 
\be
z^{-1}=\frac{1}{e^{\tau \omega N}-1}
\label{fug2}
\ee

In eq. (\ref{beta})  the quantities $\tau$ and $\omega$ 
have been related to the parameter
$h$ of the matrix model (\ref{kl1}). In the large $N$ limit these
relations simplify to
\be
{\omega}\tau \sim \frac{1}{hN} \\ \nonumber
\omega \sim 2hN.
\ee
and for $h \ll 1$ the fugacity is given by  $z\sim e^{1/h }$.
To obtain the  average particle density for $x\to 0$ we can approximate
the single particle wave function by plane waves with energy given by
$k^2$. 
\be
\rho(0) = \frac 1\pi \int_0^\infty dk \frac  1{1+z^{-1}e^{\tau k^2}}
\sim \sqrt{2}\frac{N\sqrt{h}}{\pi}
\ee
The unfolded spectral kernel is thus given by,
\be
\label{jk}
{\bar K^T}(x,0)=\frac {K^T(x  /\rho(0),0)}{\rho(0)} =
\sqrt{h}
\int_{0}^{\infty}\frac{\cos(\pi x{\sqrt{ht}})}
{2\sqrt{t}}\frac{1}{{1+z^{-1}e^t}}dt.
\ee
In the low temperature limit $h \ll 1$, 
the above spectral kernel coincides with
({\ref{ow}}) for $T = \frac {\pi h}{2}$.

 The number variance of the eigenvalues near the center of the band is
obtained by integrating the two point connected 
correlation function $R_{2,c}^T(s,0)$ including the self-correlations 
(\ref{ic}),
\be
\label{num}
\Sigma^{2}(L)= L+2\int_{0}^{L/\rho(0)}ds
(L-s)R_{2,c}^{T=\pi h/2}(s,0).
\ee
\begin{figure}[ht]
\begin{center}
\epsfig{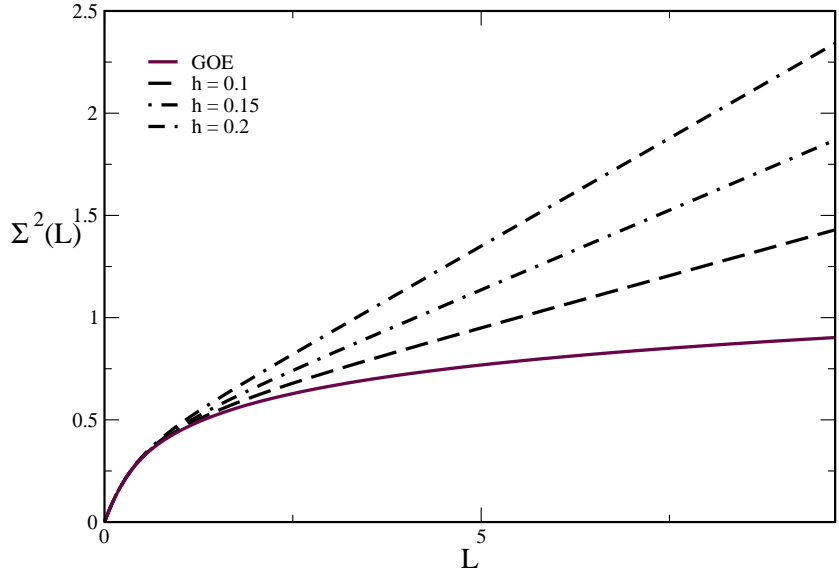}
\caption{The number variance $\Sigma^{2}(L)$ (\ref{num}) versus $L$
for $h=0.1$, $h=0.15$ and $h=0.2$.
The linear behavior of the number variance is a signature of critical
statistics. The
 slope for $h \ll 1$ is $\chi= h$.}
\label{callisto2}
\end{center}
\end{figure}
\vskip 1.5cm
The number variance $\Sigma^{2}(L)$ measures
the stiffness of the spectrum. The fluctuations are small for the GOE with
$\Sigma^{2}(L)$ proportional to $\log(L)$ for $L \gg 1$.
For the Poisson ensemble, which is an ensemble of  diagonal random 
matrices, the eigenvalues are
uncorrelated and  $\Sigma^{2}(L)=L$.
For critical statistics the number variance is asymptotically proportional
to $\chi L$.
For $\chi<<1$ the slope 
 has been connected with the multifractal dimension $D_{2}$ observed in
the  wave functions  of a disordered system undergoing a localization-
delocalization transition \cite{chalker,chalker2},
\be
\label{frac}
 \chi=\frac{d-D_{2}}{2d},
\ee
where $d$ is the spatial dimension of the system to be studied.
 As  observed in Figure 1, the number variance of our model
is linear for $L \gg1$,
with a slope $\chi = h$ for $h \ll 1$. This linear behavior
 together with the absence  
of subleading logarithmic terms in the asymptotic behavior 
of the number variance 
suggests that our matrix model describes critical statistics. 

 \section{Testing the Kravtsov-Tsvelik Conjecture}
Below we discuss two independent methods to test the Kravtsov-Tsvelik
 conjecture. 

\subsection{Conformal Calculation}

   We review  first {\cite{kawa,tsvlik} how conformal techniques can
be  utilized to calculate the
low temperature  large distance asymptotic behavior 
of the two-point correlation function of the C-S model.

 Conformal field techniques  \cite{kawa,cardy,miro1,miro} can be used to
compute the asymptotic behavior of the correlation functions of $1+1$ 
dimensional systems with a linear gapless
spectrum in the limit of large $N$ number of particles, constant
density $n=N/L$ and low temperature. 
 
  In order to identify the conformal theory associated to the low energy 
 excitations of the 1+1 dimensional  system one needs
 the value of the conformal anomaly $c$ of the
associated conformal field theory. Usually, $c$  
is obtained from the leading low
temperature behavior of the free energy of the 
system. Then, the  conformal weights
 of the primary fields of the conformal
 theory must match the leading low energy excitations 
of the $1+1$ dimensional quantum system. The latter is usually evaluated 
either numerically 
 or by finite size scaling and Bethe ansatz
techniques.
  This program was carried out for the C-S model by Kawakami and Yang
\cite{kawa}. They found that the low energy collective 
excitations of the C-S model are
described by two quantum numbers: $\Delta N$ related to excitations that
 change the number of particles and $\Delta D$ associated with excitations 
that move one particle from one Fermi point to the other. The energy, $E$, 
 and momentum, $P$, of the leading finite size excitations of the 
C-S model are given by,
 
\be
\label{1a}
E&=&\frac{2\pi }{L}\frac{\lambda}{4}\Delta D^{2}+
\frac{2\pi }{L}\frac{\lambda}{4}\Delta N^{2}+
\frac{2 \pi  (N^{+}+N^{-})}{L}, \nonumber \\
P&=&2\pi\Delta D+ P_L, \nonumber \\
P_L&=&\frac{2\pi}{L}[\Delta N\Delta D+N^{+}+N^{-}],
\ee
 where $P_L$ stands for the momentum of the finite size 
excitations and $N^{+}$ and $N^{-}$ label 
the conformal towers of states (secondary fields) 
 associated to each primary field.
 Finally, they argued, based on thermodynamics arguments, 
that the conformal anomaly associated with the C-S model is $c = 1$.
 
 In a system with conformal symmetry, 
the eigenvalues of both the Hamiltonian and 
the momentum are related to the right/left conformal 
 weights  
 $x_{n,m}/{\bar x}_{n,m}$ of the primary fields through the following relation,
\be
\label{1b}
x_{n,m}&=&\frac{L}{2\pi }[E+P_L], \nonumber\\
 {\bar x}_{n,m}&=&\frac{L}{2\pi }[E-P_L],
\ee
 where we assume that $E$ and $P_L$ depend on quantum numbers $n$ and $m$. 
Therefore, if we manage to find a conformal field theory with $c=1$ and 
eigenvalues of the momentum operator and energy operator 
 given by (\ref{1a}), the correlation functions of the C-S model
 in the asymptotic limit can be easily evaluated by means of 
conformal techniques.     
It turns out that the simplest conformal model with such 
 properties  is a  free boson 
compactified on a circle with radius $R=1/\sqrt{\lambda}$.

In general, observables do not have definite conformal dimensions  and
must be expressed as a linear combination of conformal
 excitations.  Since such conformal fields only describe 
 the excitations close to the ground state  one first has to decompose the 
 expansion of observables into ``fast'' and ``slow'' modes \cite{senec}. 
The ``slow'' modes
 are described by the conformal fields and the 
  the fast ``ones'' correspond to momenta that remain finite in the
thermodynamic limit, i.e. to excitations with $\Delta D \ne 0$.
  The density operator can be expanded as 
\be
\label{opq}
\rho(x)_{T}=\sum_{m,n=-\infty}^{\infty}c_{m,n}e^{i2\pi xn}\psi_{n,0,m}
(x)_{T}
\ee
where $\psi_{n,0,m}(x,0)_{T}$ stands for the primary state
($\Delta D=n,\Delta N=0$)  associated with the above conformal theory, 
 the phase in the expansion represents  the momentum of the ground state 
 for $L \to \infty$ (fast mode) and the index $m$ accounts for the contribution
 of secondary fields.      
 Since the density operator does not change the number of particles 
 only excitations with $\Delta N = 0$ contribute to the expansion. The
coefficients
 $c_{m,n}$ are found from the zero temperature limit ($GOE$).
 
  The density-density correlations at finite temperature 
 can be easily obtained  using
 the known result for the correlation functions
 of the conformal  fields  $\psi_{n,0,m}(x,0)_{T}$.
The first terms of the conformal prediction
for the density-density
 correlations of the C-S model at $\lambda=1$  in the low
temperature limit are thus given by
\be
\label{lc}
R_{2}^{T}(x,0)=\langle \rho(x)\rho(0)
\rangle_{T} \sim \frac{T^2}{\sinh^{2}(\pi x T)}-
\frac{T^4}{2}{\frac{\cos(2 \pi x)}{\sinh^4(\pi x T)}}-
\frac{3}{2}\frac{T^{4}}{\sinh^{4}(\pi Tx)} \ldots\, .
\ee
With a temperature as given by the Kravtsov-Tsvelik conjecture, i.e.
$T=\frac{\pi h}{2}$, the conformal result coincides with 
the  asymptotic expansion
 of the conjectured result (\ref{ic}).
Since both results have been 
obtained by using
completely different methods,
this calculation supports the validity
of the kernel (\ref{ow}) in the low temperature, long distance limit. \\

\subsection{Susceptibility}

The slope of the large distance asymptotic behavior of the number
variance is determined by the isothermal susceptibility of the C-S model
which can be obtained from the C-S partition function.
On the other hand, this slope is 
determined by an integral of the unfolded two-point cluster function
$Y_2(r)$ according to
\be
\Sigma^2(n) \sim n(1-\int_{-\infty}^\infty Y_2(r) dr).
\label{genrel}
\ee
Therefore,  agreement 
between the conformal calculation and the Kratsov-Tsvelik conjecture 
for the large distance asymptotic behavior of the two-point correlation function
does not necessarily imply that the asymptotic behavior of the number
variance is given by the susceptibility. 
 
 The susceptibility $\chi$ in the grand canonical ensemble which
measures the fluctuations of the number of
particles in a box of length $L$,
 \be
\chi = \langle N^2 \rangle- {\langle N \rangle}^2,
\ee
can be expressed as
\be
\chi = z_1 \frac d{dz_1}\langle N \rangle,
\ee
where $z_1$ stands for the fugacity and $\langle N \rangle$ is the average 
number
 of particles. Remarkably,
  the C-S gas for arbitrary statistical coupling 
$\lambda$ can still be considered a free gas but
 with exclusion statistics \cite{haldane}. Indeed,
 Sutherland
\cite{suth} has shown, by using the Bethe Ansatz 
 and a method previously developed
 by  Yang and Yang \cite{yang} for the Bose gas with a delta interaction,
 that the occupation number $n(k)$  of a gas of C-S particles
 satisfies the following transcendental equation,
\be
\label{hx}
(1-\lambda n(k)/2)^{\lambda/2}(1+(1-\lambda/2)
n(k))^{1-\lambda/2} = n(k)e^{\tau e(k) }/z_1.
\ee
For the special case $\lambda=1$, corresponding to the
 $GOE$, we obtain
\be
\label{sh}
n(k)=\frac{2}{\sqrt{1+\frac 4{z^{2}_1}e^{2\tau e(k)}}}
\ee
where $z_1$ is the fugacity for this distribution function
and $e(k) = k^2$ is the energy
 of a single particle. 
To find the relation between the fugacity and
the parameter $h$ in the Moshe-Neuberger-Shapiro model for $\lambda =1$
 we
have to use the single particle energies corresponding to the 
C-S model (\ref{er}). We thus have 
\be
N = \frac 1\pi\int_{-\infty}^\infty dx \int_{-\infty}^\infty dk
\frac{1}{\sqrt{1+\frac 4{z^{2}_1}e^{2\tau (k^2+\frac 14 \omega^2 x^2)}}}
.
\ee
The asymptotic behavior for large $z_1$ can be obtained easily by changing
to polar coordinates. This results in
\be
N = \frac{\log z^2_1}{\omega \tau}.
\ee
Using (\ref{beta}) we then find in the limit of small $h$,
\be
z_1 = e^{1/2h}.
\label{fug1}
\ee

Since the density of particles is $x$-dependent in a harmonic box
we calculate the susceptibility for particles in a rectangular box
with fugacity given by (\ref{fug1}). In this way the susceptibility
can be compared to the slope of the number variance which is 
calculated in the center of the spectrum.
 
The average number of particles in a box of length $L$ is given by
\be
\langle N\rangle = \frac L\pi \int_{-\infty}^\infty \frac {dk}
{(1+\frac 4{z^2_1} e^{2\tau k^2})^{1/2}}.
\ee
If the fugacity is parameterized as $z^2_1 \equiv e^{2\tau \bar k^2}$ we
have in the low-temperature limit,
\be
{\langle N\rangle} =
{2\bar k} \frac L\pi.
\ee
Then 
\be
\chi \equiv \langle N^2 \rangle - \langle N \rangle^2 = z_1 \frac d{dz_1}
\langle N \rangle
&=& \frac L\pi\int_{0}^\infty dk \frac {8 z^{-2}_1 e^{2\tau k^2}}
{(1+ 4 z^{-2}_1 e^{2\tau k^2})^{\frac 32}}.\nonumber 
\ee
After the change of variable $\delta k = k-\bar k$ and expanding around the 
 Fermi surface we find,
\be
\chi &=& \frac L\pi \int_{ -\infty}^{\infty} d\delta k
\frac {8e^{4\tau \bar k\delta k}}
{(1+ 4e^{4\tau \bar k \delta k})^{\frac 32}}\nn \\
&=& \frac{\langle N \rangle }{2\tau \bar k^2}=\frac{\langle N
\rangle}{2\log z_1}= h \langle N \rangle.
\label{suscep1}
\ee

The above result should be compared with the calculation of
the asymptotic behavior
 of the number variance from the two-point
spectral correlation function,
\be
\chi \sim \Sigma^{2}(\langle N \rangle)\quad {\rm for} \quad \langle
N \rangle \to \infty,
\ee
where $\Sigma^{2}(\langle N \rangle)$ is defined by
\be
\Sigma^2(\langle N \rangle ) = \langle N^2\rangle-\langle N\rangle^2 =
\langle N\rangle - 2\int_0^{\langle N\rangle}dr(\langle N\rangle -r)Y_2(r).
\ee
Here, $Y_2(r)$ is the unfolded two-point cluster function. 
According to
the Kravtsov-Tsvelik conjecture it is given by
\be
\label{i5}
Y_2(r) = K^2(r) + \frac{dK(r)}{dr} \int_r^\infty K(r') dr', 
\ee
where $K(r)$ is the kernel
\be
K(r) =\frac 1{2\pi \bar \rho} \int_{-\infty}^\infty{dk}
\frac{\cos (kr/\bar \rho)}{1+ \frac 1{z_2}e^{\tau
k^2}},
\label{kerkt}
\ee
and $\bar \rho$ is the average spectral density. 
For this cluster function, and, in fact any
cluster function that decreases stronger than $1/r$, we recover 
the relation (\ref{genrel}) for $\langle N \rangle \to \infty$. 
With the fugacity parameterized by $z_2 = e^{\tau\bar k^2}=e^{1/h}$ (see 
section 3)
 we  
have
in the low-temperature limit
\be
K(0) \equiv 1 = \frac {\bar k}{\pi \bar \rho}.
\ee
 After partial integration of the second term of (\ref{i5}) and using
that $K(0) = 1$ we obtain,
\be
\Sigma^2(\langle N \rangle) = \langle N\rangle - 2\langle N\rangle
\int_0^\infty (2K^2(r) -K(r)) dr + O(N^0).
\ee
 Integrating by parts and 
  making an  expansion about the Fermi surface $\bar k$ results in,
 
\be
K(r) =  \frac{\sin \bar k r/\bar \rho}{4\pi r}
\int_{-\infty}^\infty {ds}  \frac{\cos s r/2\tau\bar k \bar \rho}
{\cosh^2 s/2} = \frac{1}{2 \bar k \bar \rho \tau}
\frac {\sin ( \bar k  r/\bar\rho )}
{\sinh(\pi r /2 \bar k \bar \rho\tau)}.
\label{fkr}
\ee

 The integral over $r$ in (\ref{fkr}) 
can now be performed analytically resulting in the susceptibility

\be
\chi = 
\frac{\langle N \rangle}{\tau \bar k^2} 
= \frac{\langle N \rangle}{\log z_2}= h \langle N \rangle. 
\ee
This slope is in agreement 
with the result obtained from the partition function of
the C-S model. In agreement with our naive expectation, 
the value of the slope is a factor 2 larger than the one found
for the original Moshe-Neuberger-Shapiro model for 
$\lambda =2$
\cite{Moshe}.


In the kernel (\ref{kerkt}) the momentum integral
is weighted by the occupation number 
which in this case is the Fermi-Dirac distribution. 
Since the occupation number of
 the C-S model for $\lambda = 1$ is given by (\ref{sh}) it seems
more natural to make this choice instead. This results in the kernel

\be
\label{sq1}
K(x)= \frac 1{\pi\bar\rho}\int_{-\infty}^{\infty}dk\frac{\cos(2kx/\bar\rho)}
{\sqrt{1+\frac 4{z^{2}_1}e^{2\tau k^2}}}
\ee
where $z_1=e^{{\bar k}^2 \tau} \equiv e^{1/h'}$ is the fugacity. We choose
the normalization of
the kernel 
such that $K(0) =1 $. Then the  zero temperature limit ($\tau \to \infty$)
of this kernel is the usual sine-kernel, $\sin \pi x/\pi x$.
Below we show 
that the conjecture (\ref{sq1}) 
disagrees with both the conformal calculation and 
the susceptibility (\ref{suscep1}).
 
Let us  first derive the large $x$ asymptotic behavior
of the kernel (\ref{sq1}). In the low temperature limit 
 the average number of particles is again given by
$\langle  N \rangle = 2L\bar k/\pi \bar \rho$, with normalization condition
$2 \bar k /\pi \bar \rho = 1$. After partial integration the integral can
be rewritten as
\be
K(x) = \frac 1{\pi x } \int_0^\infty \frac{
\, 8 \tau k z^{-2}_1 e^{2\tau k^2} \sin( 2k x/\bar \rho )}
{(1+4z^{-2}_1e^{2\tau k^2})^{3/2}}.
\ee
In the low temperature limit the integrand is strongly peaked at
$k \approx \bar k$, and the integral can be calculated by a steepest
descent approximation
\be
K(x) &\sim& {\rm Im} \frac 2{\pi x}\int_{-\infty}^\infty du \frac{ 
4e^{2 ix (\bar k + u/\tau\bar k)/\bar \rho} e^{4u}}{(1+4 e^{4u})^{3/2}}\nn \\
   &\sim&\frac { e^{-3/2}}{3\pi} {\rm Im} \left ( \frac{ \pi i x}{\tau \bar k}
 \right )^{1/2}e^{2i\bar k x/\bar \rho - \pi x{\bar k}/2\tau 
-\pi i x (\log 4)/2\tau \bar k}.
\ee 
This asymptotic result is in disagreement with the prediction from
the conformal calculation.
 
Next we compare the asymptotic behavior of the number variance with the
susceptibility.
 The comparison of the susceptibility can again be made by 
 computing the asymptotical behavior of the number variance.
  For an exponentially decreasing kernel we
 have previously 
shown that the asymptotic behavior of the number variance is given by
\be
\Sigma_2(\langle N \rangle) = \langle N \rangle - 2\langle N \rangle
\int_0^\infty (2K^2(x) -K(x))dx + O(\langle N \rangle^0).
\label{var1}
\ee
where $K(x) $ is the kernel (\ref{sq1}) with average
eigenvalue spacing normalized to unity.
Using that
\be
\int_0^\infty \frac{\sin \pi a x}{\pi x} dx &=& \frac 12,\nn \\
\int_{-\infty}^\infty ds \frac{e^{4s}}{(1+4 e^{4s})^{3/2}} &=& \frac 18,
\ee
one easily shows that in the low temperature limit,
\be
\int_0^\infty K(x) dx  = \frac 12.
\ee
The other integral in (\ref{var1}) can be written as
\be
\int_0^\infty K^2(x) dx = \int_0^\infty dx \frac {128}{(\pi x)^2}
\int_{-\infty}^\infty dy \int_{-\infty}^y dy'
\sin((\bar k + \frac y{\tau\bar k})2 x/\bar \rho)
\sin((\bar k + \frac{y'}{\tau\bar k})2 x/\bar \rho) G(y) G(y').\nn \\
\ee
where 
\be
G(y) = \frac{e^{4y}}{(1+4 e^{4y})^{3/2}}. 
\ee
The integral over $x$ can be evaluated using the formula
\be
\int_0^\infty dx \frac {\sin(ax) \sin(bx)}{x^2} = \frac {a\pi}2
\quad {\rm for} \quad a < b.
\ee
This results in 
\be
\int_0^\infty K^2(r) dr = \frac 12 + 128
\int_{-\infty}^\infty dy \int_{-\infty}^y dy'
\frac {y'}{\pi \tau\bar k \bar \rho }    G(y) G(y').
\ee
The asymptotic behavior of the number variance is thus given by
\be
\Sigma^2(\langle N \rangle) &=& -512 \frac {\langle N \rangle}{\pi \tau\bar k 
\bar \rho} \int_{-\infty}^\infty dy \int_{-\infty}^y dy'
y' G(y) G(y')\nn \\
&=& -64 \frac {\langle N \rangle}{\pi \tau\bar k 
\bar \rho} \int_{-\infty}^\infty dy \frac {y e^{4y}}
{(1+4 e^{4y})^2} \nn \\
&=& \frac {2\langle N \rangle}{\pi \tau\bar k
\bar \rho}  \log 2\nn \\
&=& \frac {\langle N \rangle}{\log z_1
}  \log 2.
\ee
To obtain the expression after the second equality sign we have performed
a partial integration using the identity
\be
G(y)= -\frac 18 \frac d{dy}
\frac 1{\sqrt {1 + 4 e^{4y}}}.
\ee
This result for the susceptibility differs from the result  
obtained from the thermodynamic properties of the C-S gas. 
We conclude that the kernel (\ref{sq1}) does not describe the
correlations of the critical Random Matrix Model (\ref{kl1}).

\section{Critical statistics and quantum chaos}

    In this section we introduce the concept of 
 multifractal wave
functions in the context of the Anderson transition and  
 show how it may be relevant in the study 
 of deterministic quantum chaotic systems.

  By now it has been well established that  the appearance of critical 
 statistics at the Anderson transition is intimately related with 
 the multifractal properties of the wavefunctions 
{\cite{shk,chalker,chalker2,fyo}.    
We wish first to introduce  intuitively the concept of 
 multifractal wave functions \cite{kra2} . 

 Let us consider the volume of the subset of a box 
for which the absolute value of the
wave function $\Psi$ is larger than a fixed number $M$. If this volume scales
as $L^{d^*}$  (with $d^*< d$), then $d^*$ is called the fractal
dimension  $d^{*}<d$ of $\Psi$. 
In case  the fractal dimension depends on 
the value of $M$, the wave function is said to be multifractal.
More formally, multifractality is defined through the inverse 
participation ratio,
\be
I_{p}=\sum_{r}\langle|\Psi_{n}(r)|^{2p}\rangle \propto L^{-D_{p}(p-1)}
\ee
where $\Psi_{n}$ is the wave function with energy $E_{n}$, and $D_{p}<d$ 
is a set of exponents characterizing the anomalous (multifractal) 
scaling of the moments of the wave function. We remark that, although  
confined to  fractal subsets of the sample, wave functions  of such
systems  overlap strongly when their energies are close enough \cite{fyo}. 
Such strong overlap is responsible for the short-range level 
repulsion observed at the Anderson transition.      
It is worthwhile to note  that this anomalous scaling has, in principle, 
a pure quantum mechanical origin.   As the density of impurities 
 increases, the deBroglie wavelength of the particles  
becomes comparable with the mean free path and localization 
effects start to be relevant. 
We stress that the classical dynamics of the Anderson transition  
does not provide us with valuable information 
to describe quantum spectral correlations.
One may wonder to what extent such multifractal behavior may be observed
in deterministic quantum chaotic systems. 

What has become known as the Bohigas-Giannoni-Schmit conjecture
\cite{BGS} is that generically quantum spectra of 
classically chaotic systems are correlated according to
the Wigner-Dyson random matrix ensembles, whereas
spectral correlations of classically integrable  systems are 
close to Poisson statistics. 
In most cases, by modifying the parameters of the system,
a transition   from integrable to chaotic dynamics can be observed. 
If the KAM theorem is applicable, 
this transition is smooth and both integrable and chaotic regions 
coexist until the last KAM torus is completely destroyed. 
Although spectral statistics of such mixed system have been described in 
terms of banded random matrix models \cite{SVZ}, they are believed
to be non-generic and different form critical statistics \cite{guhr}. 

\begin{figure}[ht]  
\begin{center}
\epsfig{figure=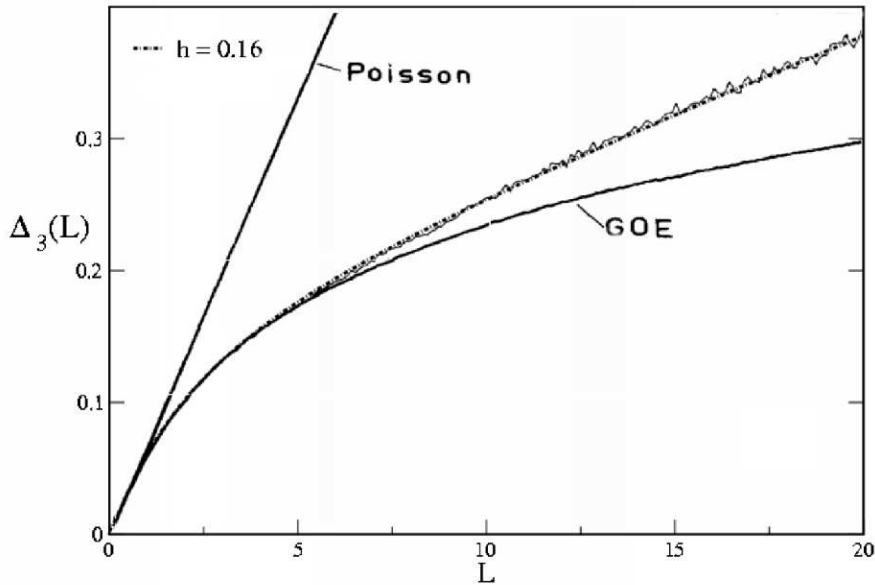,width=130mm,angle=0}
\caption{The spectral rigidity $\Delta_{3}$  of the 
 anisotropic Kepler problem obtained in \cite{marxer} versus
the prediction 
 of our model for $h=0.16$ obtained from (\ref{jk}). 
Both curves are barely distinguishable. The
 numerical data are reprinted from Fig 2. in Ref. \cite{marxer}}
\label{callisto6}
\end{center}
\end{figure} 
\vskip 1.5 cm
The situation is different in cases where 
 the KAM theorem does not apply. 
In those systems, the invariant KAM curves may not exist 
 at all and small changes in the coupling constant 
 can produce qualitative modifications in the classical phase space.
 The dynamics is, in general, intermediate between chaotic and integrable. 
 The lack of KAM tori permits a particle explore the full available classical 
 phase space without having full chaotic motion. 
 Such forms of phase space are also known as
 stochastic webs \cite{zaslav}. 
In certain cases, the classical  
 phase space becomes increasingly intricate, showing both 
self-similar and fractal properties \cite{zaslav}. 
We notice that such a structure is reminiscent 
 of the way that the KAM tori break up into  
"fractal" orbits of zero dimension \cite{bak} (cantori) 
 as the system becomes chaotic.
 Classically, cantori represent strong obstacles to phase space transport. 
 Our aim is to study the effect, if any, 
of such self-similar structure in the spectral correlations of the 
quantum counterpart.
  Roughly speaking, the influence of cantori on the quantum dynamics 
will depend on the relation between  the size of the cantori 
and   Planck's constant. 
For cantori smaller than the Planck cell, quantum dynamics 
cannot resolve the
classical fine structure. In this case, cantori act 
as perfect barriers to the quantum motion resembling the effect 
of a classically integrable system and 
the spectral correlations of the 
quantum counterpart are close to Poisson statistics.
 In the intermediate case the situation is less clear. Recently,
it has been reported \cite{casat} that cantori drive spectral 
correlations smoothly 
 from Poisson to RMT as the system approaches to the ergodic regime. 
   
   Below, we present numerical evidence 
 that deviations from GOE statistics 
 caused by the self similar structure of the classical phase space
 may be described
 by critical statistics at least while the deviations from GOE are small.

\subsection{The Anisotropic Kepler problem}

The anisotropic Kepler Hamiltonian
\be
H=\frac{1}{2}p^{2}_{\rho}+\frac{1}{2}\gamma p_{z}^{2}-\frac{1}{r}
\ee
 is an interesting example of a non-KAM system undergoing an abrupt chaotic
 integrable transition. It has been utilized as a model of
 donor impurities in a semiconductor \cite{kohn,rod}.
  Even for small departures from the integrable case, $\gamma =1$,
 the classical phase space is densely filled with remnants of cantori
 \cite{marxer}.  
 Gutzwiler has shown that for $\gamma < 8/9$ the orbits in phase space
 can be uniquely represented in terms of symbolic dynamics. Such
  representation is a signature of hard chaos. Indeed, for
  $\gamma < 1/2$ there are no islands of stability in phase space.
Furthermore,   
the measure of the surface of section based on the symbolic 
dynamics is multifractal with respect to the usual 
 Liouville measure.  
  Since the periodic orbits can be effectively enumerated,
 the energy levels of the quantum counterpart can be 
approximately evaluated by means 
of analytical techniques \cite{bogomol}.   
 
A numerical 
 study of the spectral correlations of highly excited 
 states of this system was carried out in 
\cite{marxer}. In a basis in which 
the  Hamiltonian has a band structure, they succeeded to obtain
up to $5500$ energy levels. In Fig. 2 we show their result
for the spectral rigidity of the spectrum from level 
$2501$ to $5500$. 
  As observed, the deviations 
  from the GOE are very well 
described by the
 critical random matrix model (\ref{kl1}). 
Based on the analogy 
 with disordered systems, we conjecture
 that the wavefunctions of this system are multifractal. We are not aware
of numerical results that can confirm or disprove this conjecture.
\begin{figure}[!ht]  
\begin{center}
\epsfig{figure=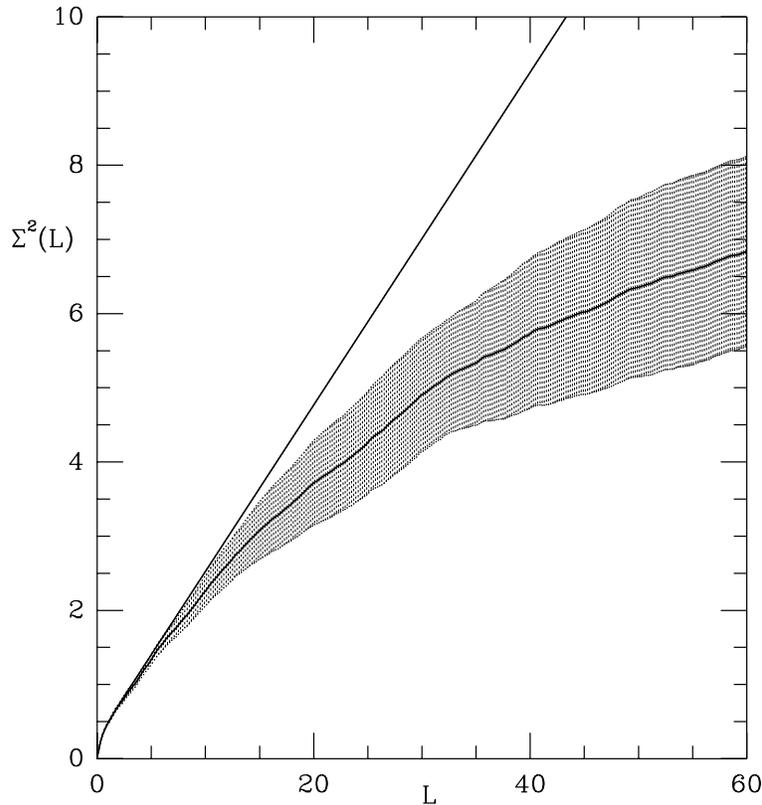,width=100mm,angle=0}
\caption{The number variance $\Sigma^{2}(L)$ versus $L$. 
The number variance of the quasi-energy levels of a kicked particle  
in an infinite potential well at $K=kT=50$ \cite{hu}  agrees 
with the prediction  of the critical GOE (\ref{num}) for $h=0.2122$ (upper curve) up to
 $10$ eigenvalues. 
The downward tendency of the numerical result may be due to  
finite size effects. 
The error in the numerical results is indicated by the thickness of the curve.
We thank to Baowen Li \cite{hu} 
for  kindly providing us with 1024 energy levels to compute the 
 number variance.}
\label{callisto5}
\end{center}
\end{figure} 
\vskip 1.5cm  
\subsection{ Kicked particle in a infinite potential well}
Recently, in {\cite{hu}},  another non-KAM system with similar properties, 
a kicked particle in a infinite 
potential well, was studied both quantum mechanically 
 and classically. The Hamiltonian is given by,
\be
\label{i}
H=\frac{p^2}{2}+V(x)+k\cos(x+1)\sum_{n=-\infty}^{+\infty}\delta(t-nT)
\ee
where $V(x)$ is an infinite well potential of length $\pi$, $T$ 
is the period of the kick and $k$ the strength.
Concerning the classical motion, the KAM theorem is not 
 applicable because the potential is not smooth. Indeed, it was found 
 \cite{hu} that  
 the classical phase space resembles a stochastic web  
 with a self similar structure. This is in contrast 
with the standard kicked rotor
 where the classical phase space is a mixture of chaotic an integrable parts
 separated by KAM tori.  

 The quantum mechanical properties of the model are described by the evolution
 operator  ${\hat U}=e^{-i{\hat p}^2T/4}e^{k\cos(x+1)}e^{-i{\hat p}^2T/4}$ 
over a period $T$ of the kick.
 The quasi energies associated to this operator were obtained in {\cite{hu}}
 by diagonalizing $\hat U$ in a basis of $1024$ eigenstates of 
the free Hamiltonian $\langle q|n\rangle=\sqrt{\frac 2{\pi}}\sin(nq)$.   

 Unlike the kicked rotor where
  the matrix evolution has an exponential decay in a basis of plane waves, 
 it can be shown that the matrix elements 
  of $ \hat U$ are well described by a random banded
 matrix (RBM) with power-like decay,
 $|\langle m|\hat U|m+n\rangle| 
\propto \frac{b^2}{n^2}$ for $b \ll n$ and constant for $b \gg n$ 
where $b \sim K = kT$ is the size of the band.  
 
In agreement with the results of {\cite{dittes}}, the nearest 
neighbor distribution  reported in {\cite{hu} smoothly interpolates  
between Poisson and $GOE$ as $K$ is increased. Recently, an experimental
 realization of this model  was studied in \cite{from3}. \\
\begin{figure}[!ht]  
\begin{center}
\epsfig{figure=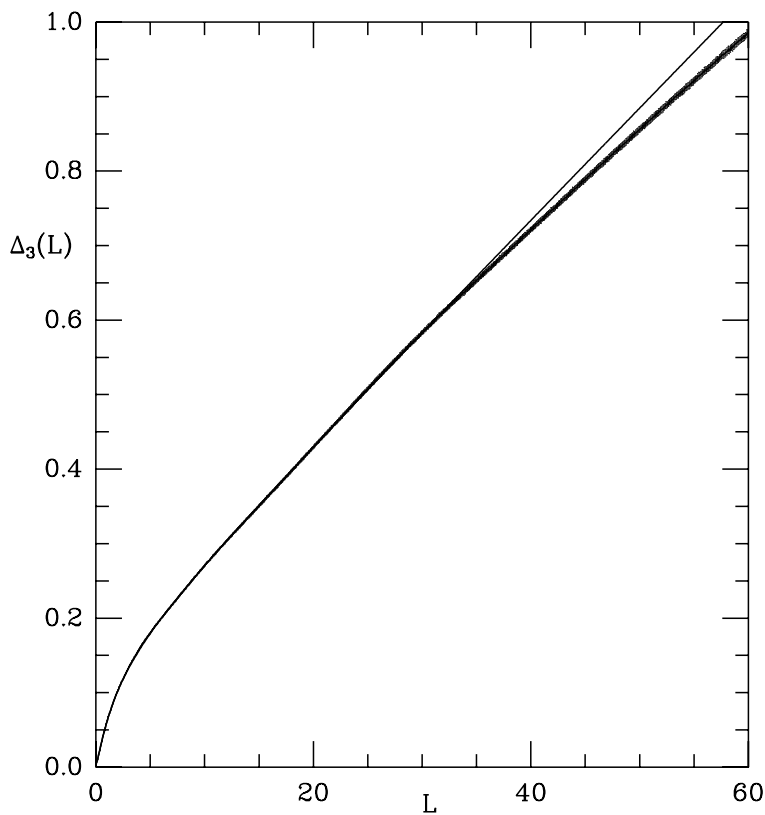,width=100mm,angle=0}
\caption{ The spectral rigidity $\Delta_{3}(L)$ obtained in 
\cite{hu} for the energy levels 
of a kicked particle in an infinite potential well at $K=kT=50$ is 
 accurately described by the critical kernel (\ref{jk})
 at $h=0.2122$ (upper curve). The error in the numerical results 
is indicated by the thickness of the curve.}
\label{callisto4}
\end{center}
\end{figure}

{\bf Analysis of results}\\
In Figures 3 and 4, we show the number variance and the $\Delta_3$-statistic
of the sequence of 1024 eigenvalues obtained in \cite{hu} for $kT = 50$.
The upper curve is the analytical result derived from the
two-point correlation function (\ref{ic}) with kernel (\ref{jk}). In both 
 cases the  
 numerical result is plotted with its error (see below). 

For the $\Delta_3$-statistic, the $\chi^2 $ 
is minimized on the interval $[0,30]$ for $h = 0.2122$ with a 
value\footnote{Here and below we find values of $\chi^2$ that are
significantly less than one. This is possible because 
the values of $\Delta_3(L)$ for different values of $L$ are correlated.
Therefore, our values of $\chi^2$ have to be used with care and 
cannot be interpreted in terms of a $\chi^2$ distribution.}
 of $\chi^2 = 0.32$. 
The errors in the definition of 
 $\chi^2$ have been calculated by splitting the 1024 eigenvalues 
into eight ensembles of 128 eigenvalues and evaluating the number variance 
for each ensemble separately (denoted by $\Sigma^2_i(L), i = 1, \cdots, 8$).
The error in the number variance is thus given by
\be
\sigma(L) = \frac 1{\sqrt 8}
\left [\frac 18\sum_{i=1}^{8}(\Sigma_{i}^2(L)-\Sigma_{\rm mean}^2(L))^2
\right ] ^{1/2}\quad {\rm where} \quad
\Sigma_{\rm mean}(L)=\frac 18 \sum_{i=1}^{8}\Sigma_i^2(L).
\ee
The error in the $\Delta_3$-statistic is obtained from this error by means
of a Monte-Carlo simulation using the relation between $\Delta_3(L)$ and
$\Sigma^2(L)$ \cite{mehta}. 
As observed in Figure 4, the error in $\Delta_3(L)$ is much 
 smaller that the error in $\Sigma^{2}(L)$.

 Next we ask the question whether 
the  asymptotic behavior of the spectral rigidity 
 is  linear without a logarithmic correction, just
as in the analytical case. 
If a logarithmic correction is absent,
one can almost discard the possibility of a mixed classical phase space
  as the reason of the observed deviation from the GOE. 
 In order to prove the absence of such term we fit the numerical
 curve ${\Delta_{3}(L)}_{{\rm num}}$ to 
${\Delta_{3}(L)}_{{\rm fit}}= a+bL+c\log L$.
  For instance, on a interval
 $[15,25]$ a best fit is obtained for $a = 0.110\pm0.004$ , $b = 0.0156\pm 0.0001$, $c = 0.002\pm 0.002$ with a value of $\chi^2 = 0.015$. We find that the 
value of the coefficient $c$ is compatible with zero. This suggests that the classical phase space is not a mixture of chaotic and integrable regions.

 In the case of the number variance, because of the size of the error, 
no conclusive evidence on the absence of the logarithmic term can be obtained 
by such fit.
  At large distances 
the number variance seems to deviate from a linear behavior 
by a quadratic term. Although such terms are typically 
caused by finite size effects, 
we do not have a clear understanding of its origin. Since 
the $\Delta_3$-statistic projects out a quadratic dependence
  of the number variance the linear behavior persists 
to much larger distances in this case (see Figure 4).
    
 Finally, let us confront the conjecture (\ref{ic}) with our present numerical
results. As we mentioned 
 previously, it may look reasonable to replace the kernel (\ref{jk}) by 
 (\ref{sq1}).  It can be shown 
 that  the spectral rigidity obtained from the kernel (\ref{sq1}) is almost 
 indistinguishable from the one obtained 
 from (\ref{jk}) and therefore the agreement with 
 the numerical result is expected to be equally good with one fitting
parameter at our disposal.
However, a more careful analysis shows that the kernel (\ref{sq1}) leads
to a value of $\chi^2$
much higher than the one obtained from (\ref{jk}). 
  For the interval $[0,30]$, using the kernels
(\ref{jk}) and (\ref{sq1}), a best fit is obtained for $h = 0.2122$ with 
a value of  $\chi^{2}_{\rm Fermi} = 0.32 $ and for $h' = 0.325$ with a value
 of $\chi^{2}_{\rm Sqrt}= 2.27$,  respectively.

The above findings suggest
 that a  self-similar classical phase space dominated by cantori  
 has a strong impact on the quantum spectral correlations. Critical 
statistics appears as the leading candidate to describe 
 such correlations and, consequently, enlarge the range of applicability 
 of random matrix ensembles.  
 
Finally, we list  other quantum systems between integrable and chaotic whose 
 quantum 
spectral correlations show a similarity  with critical statistics:   
 quantum billiards with a point 
scatterer  
\cite{arve,weav}, the Kepler billiard \cite{bogo3,levitov}, 
semiconductor billiards \cite{ketz,from}, 
the stadium billiard inside certain range of parameters 
\cite{shudo}. For applications 
concerning pseudo integrable billiards we refer to \cite{bogo3}.  

\section{Conclusions}

In this article we have introduced
a one parameter ensemble of symmetric random matrices.
 This ensemble  interpolates  between  
 the Gaussian Orthogonal Ensemble and the Poisson ensemble and is capable
 of describing critical statistics.

  We have shown that, in an eigenvalue basis, 
the joint eigenvalue distribution of our model
 coincides with the diagonal density matrix of the C-S model at finite
 temperature where the additional parameter 
 of  the matrix model plays the role of temperature. 
Remarkably, this equivalence can be extended to all random 
matrix ensembles 
associated with the large families of symmetric 
 spaces according to the Cartan classification 
  thus providing a novel link between strongly interacting quantum 
systems and random
  matrix theory. 
     
 We have calculated the spectral correlation functions based on 
a recent conjecture  by Kravtsov and Tsvelik
for the correlation functions  of the C-S model
in the low temperature limit. 
We have tested the validity
 of this conjecture by two
  independent methods:  one based on the effective conformal symmetry
 of the C-S model in the low temperature limit and the other based on  the
thermodynamical properties of a gas of particles  governed by
the C-S Hamiltonian. We have found that both the long distance low
temperature
behavior of the two-point correlation function obtained from the conformal
calculation and the susceptibility of the C-S model agree with the
conjecture made by Kravtsov and
Tsvelik. 
         
Based on  the Kravtsov-Tsvelik conjecture we
find that, although level repulsion is still present,
the number variance is asymptotically linear with a slope less than one
and no subleading logarithmic term present. This indicates  
that our random matrix model describes critical statistics. 

 Finally, we have argued 
 that critical statistics is relevant to describe
 spectral correlations of chaotic quantum systems 
 for which the Poincare section of 
 the classical counterpart is globally self-similar/fractal.
  Two examples with such classical phase space, 
a kicked particle in a potential well and the anisotropic 
 Kepler problem, have been discussed in detail.
   In both cases, long range spectral 
correlators such as the 
 number variance  and the $\Delta_3$-statistic 
are accurately described by our analytical results based on the 
Kravtsov-Tsvelik conjecture.
Indeed, for the kicked particle, we have shown
 that the spectral rigidity is asymptotically linear with no subleading 
 logarithmic term present.  This may  be an indication that the 
wavefunctions of this model show multifractal properties.

 \vskip 0.5 cm
{\bf Acknowledgements}
We thank Denis Dalmazi, Shinsuke Nishigaki and Vladimir Kravtsov 
 for important suggestions and useful 
discussions. We are indebted to Baowen Li for 
providing us with their numerical data. 
This work was supported by the US DOE grant DE-FG-88ER40388.

\end{document}